\documentclass[a4paper, 10pt,onecolumn]{article} 
\usepackage{times}
\usepackage{amssymb}
\usepackage{amsmath}
\usepackage{psfrag}
\usepackage{graphicx} 

\newtheorem{proposition}{Proposition}

\date{February 2008}
\begin{document}

\title{Design and performance evaluation of a state-space based AQM}

\author{Yassine Ariba$^{\ast\dag}$, Yann Labit$^{\ast\dag}$ and Fr\'ed\'eric Gouaisbaut$^{\ast}$
\thanks{ Universit\'e de Toulouse; UPS, 118 Route de Narbonne, F-31062 Toulouse, France.}%
\thanks{ LAAS; CNRS;  7, avenue du Colonel Roche, F-31077 Toulouse, France.
	Email: {\tt\small\{yariba, ylabit, fgouaisb\}@laas.fr}}
}

\maketitle
\thispagestyle{empty}

\begin{abstract}
Recent research has shown the link between congestion control in communication networks and feedback control system. In this paper, the design of an active queue management (AQM) which can be viewed as a controller, is considered. Based on a state space representation of a linearized fluid flow model of TCP, the AQM design is converted to a state feedback synthesis problem for time delay systems. Finally, an example
extracted from the literature and simulations via a network simulator NS (under cross traffic conditions)
support our study. 
\end{abstract}

\section{Introduction}

Congestion control is a very active research area in network community. In order to supply the well known transmission control protocol (TCP), active queue management mechanisms have been developed. AQM regulates the queue length of a router by
actively dropping packets. Various
mechanisms have been proposed in the literature such as
Random Early Detection (RED) \cite{Flo93}, Random Early Marking
(REM) \cite{Ath00}, Adaptive Virtual Queue (AVQ) \cite{Kun01}
and many others \cite{Ryu04}. Their performances have been evaluated in \cite{Ryu04}
and empirical studies have shown their effectiveness (see \cite{Le03}). Recently, significant studies proposed by \cite{Hol02} have redesigned the AQMs using control theory and $P$, $PI$ have been developed in order to cope with the packet dropping problem. Then, using dynamical model developed by \cite{Mis00}, many research have been devoted to deal with congestion problem in a control theory framework (for example see \cite{Tar05}). Nevertheless, most
of these papers do not take into account the delay and ensure the
stability in closed-loop for all possible delays which could be
conservative in practice.\\
Modeling the congestion control using time delay is not new and global stability analysis has been studied by \cite{Mic06} and \cite{Pap04} via Lyapunov-Krasovskii theory. Also, in \cite{Kim06}, a delay dependent state feedback controller is provided by
compensation of the delay with a memory feedback control. This latter methodology is interesting in theory but hardly suitable
in practice.\\
 Based on a recently developed Lyapunov-Krasovskii
functional, an AQM stabilizing
the TCP model is designed. This synthesis problem is carried out as state feedback synthesis for time delay systems. Then, this method is applied on an augmented system in order to vanish the steady state error in spite of disturbance.
\\ The paper is organized as
follows. The second part presents the model
of a network supporting TCP and the time delay system representation. Section \ref{synthesis} is dedicated to the
design of the AQM ensuring the stabilization of TCP. Section \ref{simulation} presents application
of the exposed theory and simulation results using NS-2 (see \cite{Fal02}) before concluding this work. 
\section{Problem statement}
\label{probleme}
\subsection{The linearized TCP fluid-flow model}
\begin{figure}
\begin{center}
\vspace*{-1em}
\includegraphics[height=4cm,width=8cm]{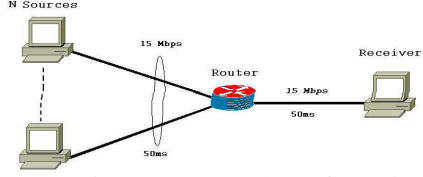}
\vspace*{-4.5em}
\caption{Network configuration}
\label{topology}
\end{center}
\end{figure}

\begin{figure}
\begin{center}
 \psfrag{p0}[][][0.7]{$p_0$}
 \psfrag{W0}[][][0.7]{$W_0$}
 \psfrag{deltaq}[][][0.7]{$\delta q(t)$}
\includegraphics[height=3cm,width= 5.5cm]{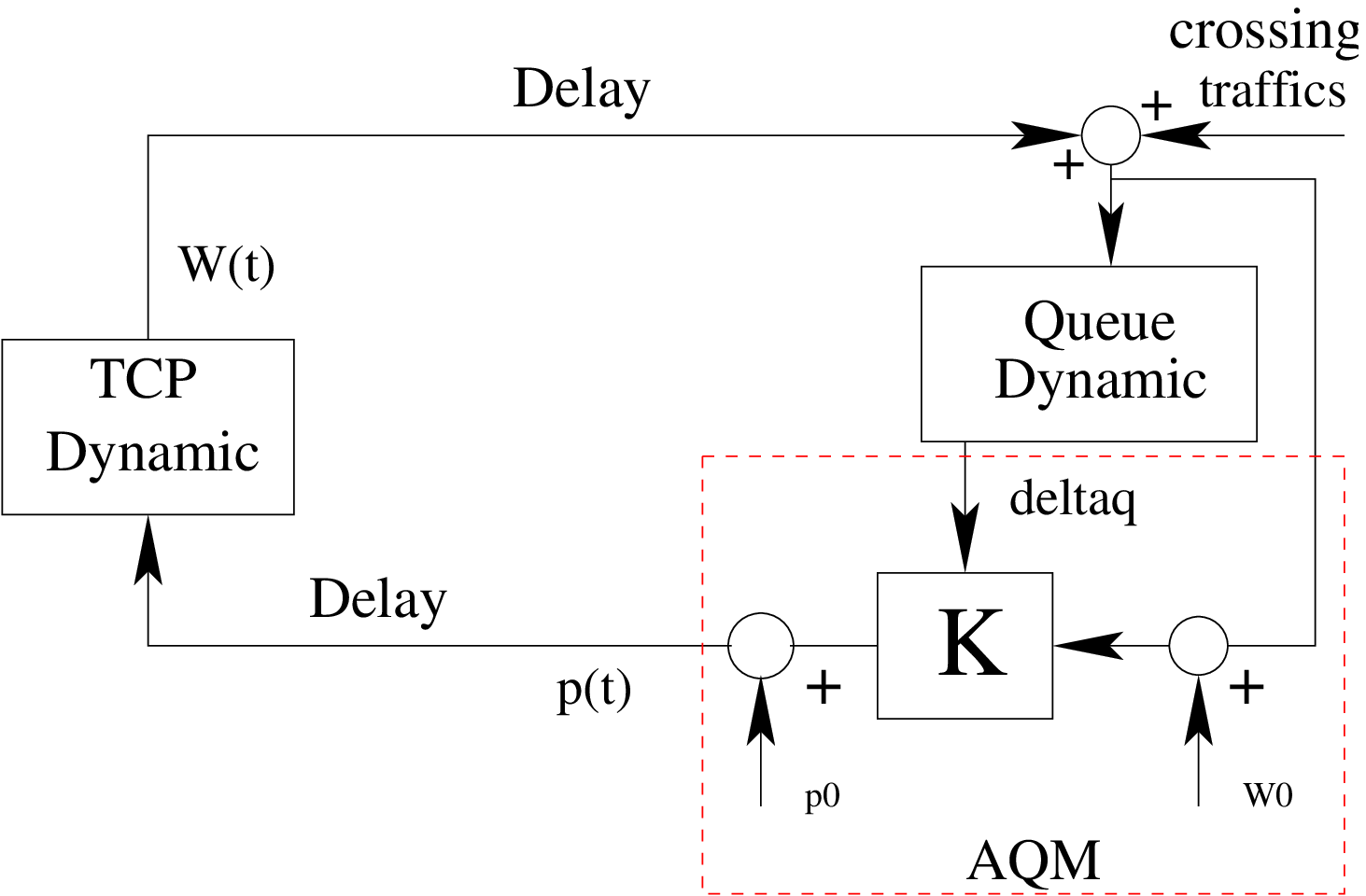}
\caption{Design of an AQM as a state feedback}
\label{schema_SF}
\end{center}
\end{figure}

In this paper, we consider the network topology consisting of $N$ homogeneous TCP sources (i.e with the same propagation delay) connected to a destination node through a router (see figure \ref{topology}).
The bottleneck link is shared by $N$ flows and TCP applies the well known congestion avoidance algorithm to cope with the phenomenon of congestion collapse \cite{Jac88}. Many studies have been dedicated to the modeling of TCP and its AIMD (additive-increase multiplicative-deacrease) behavior \cite{Low02}, \cite{Sri04}, \cite{Tar05} and references therein. We consider in this note the model (\ref{modelNL}) developed by \cite{Mis00}. This latter may not capture with high accuracy the dynamic behavior of TCP but its simplicity allows us to apply our methodology. Let consider the following model
\begin{equation}
  \label{modelNL}
  \left\{\begin{array}{rcl}
\dot{W}(t)&=&\frac{1}{R(t)}-\frac{W(t)W(t-R(t))}{2R(t-R(t))}p(t-R(t))\\
\dot{q}(t)&=&\frac{W(t)}{R(t)}N-C+d(t)
\end{array} \right.
\end{equation}
where $W$ is the TCP window size, $q$ is the queue length of the router buffer, $R$ is the round trip time (RTT) and can be expressed as $R=q/C+T_p$. $C$, $T_p$ and $N$ are parameters related to the network configuration and represent the transmission capacity of the router, the propagation delay and the number of TCP sessions respectively. The variable $p$ is the marking/dropping probability of a packet (that depends whether the ECN option, explicit congestion notification, is enabled, see \cite{Ram99}).
In the mathematical model (\ref{modelNL}), we have introduced an additional signal $d(t)$ which models cross traffics through the router and filling the buffer. These traffics are not TCP based flows (not modeled in TCP dynamic) and can be viewed as perturbations since they are not reactive to packets dropping (for example, UDP based traffic).
 A linearization and some simplifications of (\ref{modelNL}) was carried out in \cite{Hol02} to allow the use of traditional control theory approach. The linearized fluid-flow model of TCP is as follow,
\begin{equation}
  \label{modelL}
    \left\{\begin{array}{ll}
\delta\dot{W}(t)=-\frac{N}{R_0^2C}\Big(\delta W(t)+\delta W(t-h(t))\Big)\\ \hspace*{0.7cm}-\frac{1}{R_0^2C}\Big(\delta q(t)-\delta q(t-h(t))\Big)-\frac{R_0C^2}{2N^2}\delta p(t-h(t))\\
\delta\dot{q}(t)=\frac{N}{R_0}\delta
W(t)-\frac{1}{R_0}\delta q(t)+d(t)
\end{array}\right.
\end{equation}
where $\delta W \doteq W-W_0$, $\delta q \doteq q-q_0$ and $\delta p
\doteq p-p_0$ are the perturbated variables
about the operating point. The operating point $(W_0,q_0,p_0)$ is defined by
\begin{equation*}
  \label{point_eq}
    \left\{\begin{array}{l}
\dot{W}=0~\Rightarrow~W_0^2p_0=2\\
\dot{q}=0~\Rightarrow~W_0=\frac{R_0C}{N},~R_0=\frac{q_0}{C}+T_p
\end{array}\right.
\end{equation*}
The input of the model (\ref{modelL}) corresponds to the drop probability of a packet. This probability is fixed by the AQM. This latter has for objective to regulate the queue size of the router buffer.
In this paper, this regulation problem is addressed in Section \ref{synthesis} with the design of a stabilizing state feedback for time delay systems. Indeed, an AQM acts as a controller (see figure \ref{schema_SF}) and in order to design it, we have to solve a synthesis problem. Considering a state feedback, the queue management strategy of the drop probability will be expressed as
\begin{equation}
  \label{SF}
  p(t)=p_0+k_1\delta W(t)+k_2\delta q(t).
\end{equation}
where $k_1$ and $k_2$ are the components of the matrix gain $K$ which we have to design. Note that the input $p(t)=u(t)+p_0$ of the system (\ref{forme_can}) is delayed.

\subsection{Time delay system approach}
In this paper, we choose to model the dynamics of the queue and the
congestion window as a time delay system. Indeed, the delay is an
intrinsic phenomenon in networks and taking into account its
characteristic should improve the precision of our model with respect to the TCP behavior.
\\ The linearized TCP fluid model (\ref{modelL})
can be rewritten as the following time delay system:
\begin{equation}
\label{forme_can} \left\{\begin{array}{l}
\dot{x}(t)=Ax(t)+A_dx(t-h)+Bu(t-h)+B_dd(t)\\
x_0(\theta)=\phi(\theta),~\mbox{with }\theta\in[-h,0]\end{array}\right.
\end{equation}
with
\begin{equation}
\label{esp_etat}
{\small A\!=\!\left[\begin{array}{cc}\! -\frac{N}{R_{0}^{2}C}\! &\!
-\frac{1}{CR_{0}^2}\! \\\! \frac{N}{R_0}\! &\! -\frac{1}{R_0}\!
\end{array}\right]\!,\!~A_d\!=\!\left[\begin{array}{cc}\! -\frac{N}{R_{0}^{2}C}\! &\!
\frac{1}{R_{0}^2C}\! \\\! 0\! &\! 0\! \end{array}\right]\!,\!B\!=\!\left[\begin{array}{c}\!
-\frac{C^{2}R_0}{2N^2}\!\\\! 0\! \end{array}\right]\!}
\end{equation}
$B_d=[0~~1]^T$, $x(t)=[\delta W(t) ~~\delta
q(t)]^T$ is the state vector and $u(t)=\delta p(t)$ the input.
$\phi(\theta)$ is the initial condition.\\ There are mainly
three methods to study time delay system stability: analysis of the
characteristic roots, robust approach and Lyapunov theory. The
latter will be considered because it is an effective and practical
method which provides LMI and BMI (Linear/Bilinear Matrix Inequalities,
\cite{Boy94}) criteria. To analyze and control the system
(\ref{forme_can}), the Lyapunov-Krasovskii approach (see \cite{Gu03}) is
used which is an extension of the traditional Lyapunov
theory.

\section{Stabilization: design of an AQM}
\label{synthesis}
In Section \ref{probleme}, the model of TCP/AQM has been addressed as time delay system. The congestion problem needs the construction of a controller which regulates the buffer queue length.
In this section, we are first going to present a delay dependent stability analysis condition for time delay systems. Then, based on this criterion, a synthesis method to derive a stabilizing state feedback is deduced.

\subsection{Stability analysis of time delay systems}

In this subsection, our goal is to derived a condition which takes into account an upperbound of the delay. The delay dependent case starts from a system stable without delays and looks for the maximal delay that preserves stability.\\
 Usually, all methods involve a Lyapunov functional,
and more or less tight techniques to bound some cross terms and to
transform system \cite{Gu03}. These choices of specific
Lyapunov functionals and overbounding techniques are the origin of
conservatism. In the present paper, we choose a recently developed
Lyapunov-Krasovskii functional (\ref{lyap_ddr}) \cite{Gou06b}:
\begin{equation}
\label{lyap_ddr} \begin{array}{c}
 V(x_t)=x^T(t)Px(t)+\int\limits_{t-\frac{h}{r}}^{t}\!\!\int\limits_{\theta}^{t}\dot{x}^T(s)R\dot{x}(s)dsd\theta\\[0.5em] ~~~~~+\int_{t-\frac{h}{r}}^{t}\left(\begin{array}{c} x(s)\\x(s-\frac{1}{r}h)\\\vdots\\x(s-\frac{r-1}{r}h)\end{array}\right)^TQ\left(\begin{array}{c} x(s)\\x(s-\frac{1}{r}h)\\\vdots\\x(s-\frac{r-1}{r}h)\end{array}\right)ds
\end{array}
\end{equation}
 where $P\in\sf{S}^n$ is a positive definite matrix,
$Q\in\sf{S}^{rn}$ and $R\in\sf{S}^n$ are two positive definite
matrices. $r\geq 1$ is an integer corresponding to the discretization
step. Using this functional, Let us introduce the following proposition.
 \begin{proposition}
\label{prop_dd} If there exist symmetric positive definite matrices
$P$, $R\in\sf{R}^{n\times n}$, $Q\in\sf{R}^{rn\times rn}$, a scalar $h_{m}>0$ and an integer $r\geq 1$
 such that
\begin{equation}
\label{analyse_dd}  S^{\perp^T}\Gamma S^\perp< \sf{0}
\end{equation}
 where
\begin{equation}
\label{gamma} {\tiny \Gamma=\!\left[\begin{array}{ccccc} \frac{h_{m}}{r}{\bf
R}&{\bf
P}&\sf{0}&\ldots&\sf{0}\\{\bf
P}&-\frac{r}{h_{m}}{\bf R}&\frac{r}{h_{m}}{\bf
R}&~&\vdots\\\sf{0}&\frac{r}{h_{m}}{\bf
R}&-\frac{r}{h_{m}}{\bf
R}&~&\vdots\\\vdots&~&~&\ddots&\vdots\\\sf{0}&\ldots&\ldots&\ldots&\sf{0}
\end{array}\right]+\left[\begin{array}{ccc}
\sf{0}&\ldots&\sf{0}\\\vdots&{\bf
Q}&\vdots\\\sf{0}&\ldots&\sf{0}
\end{array}\right]+\left[\begin{array}{cc}
\sf{0}&\ldots\\\sf{0}&\ldots\\\vdots&{\bf
Q} \end{array}\right]}
\end{equation}
 and
\begin{equation}
\label{defS}
S=\left[\begin{array}{cccc} -\sf{1} & A &\sf{0}_{n\times (r-1)n}&A_d \end{array}\right]
\end{equation}
then, system (\ref{forme_can}) (with $u(t)=0$ and $d(t)=0$) is stable for all $h\leq
h_{m}$. \end{proposition}
~\\ \underline{\it Proof}: It is always possible to rewrite
(\ref{forme_can}) as $S\xi=\sf{0}$ where
{\small \begin{equation}
   \label{proj_ddr}
\begin{array}{c}
   \xi= \left[\begin{array}{c} \dot{x}(t)\\x(t)\\x(t-\frac{1}{r}h)\\ \vdots\\x(t-\frac{r-1}{r}h)\\x(t-h) \end{array}\right]\in\sf{R}^{(r+2)n}
\end{array}
 \end{equation}}
and $S$ is defined as (\ref{defS}).
 Using the extended variable $\xi(t)$ (\ref{proj_ddr}), the
derivative of $V$ along the trajectories of system (\ref{forme_can})
leads to:
 \begin{equation}
 \left\{\begin{array}{rcl}
   \dot{V}(x_t)&=&\xi^T \left[\begin{array}{ccccc} \frac{h}{r}{\bf R}&{\bf P}&\sf{0}&\ldots&\sf{0}\\{\bf P}&-\frac{r}{h}{\bf R}&\frac{r}{h}{\bf R}&~&\vdots\\\sf{0}&\frac{r}{h}{\bf R}&-\frac{r}{h}{\bf R}&~&\vdots\\\vdots&~&~&\ddots&\vdots\\\sf{0}&\ldots&\ldots&\ldots&\sf{0} \end{array}\right]\xi\\[0.5em] &+&\xi^T \left[\begin{array}{ccc} \sf{0}&\ldots&\sf{0}\\\vdots&{\bf Q}&\vdots\\\sf{0}&\ldots&\sf{0} \end{array}\right]\xi-\xi^T \left[\begin{array}{cc} \sf{0}&\ldots\\\sf{0}&\ldots\\\vdots&{\bf Q} \end{array}\right]\xi < \sf{0}\\[0.5em]
 \mbox{s.t.}&& \left[\begin{array}{cccccc} -\sf{1}&A&\sf{0}&\cdots&\sf{0}&A_d \end{array}\right] \xi=0
 \end{array} \right.
 \end{equation}
\begin{equation}
   \label{proj_ddr2}
\Leftrightarrow \left\{\begin{array}{ll}
  \dot{V}(x_t)=\xi^T\Gamma\xi< \sf{0}\\
 \mbox{s.t. } \left[\begin{array}{cccccc} -\sf{1}&A&\sf{0}&\cdots&\sf{0}&A_d \end{array}\right] \xi=0
 \end{array} \right.
\end{equation}
 where $\Gamma\in\sf{S}^{(r+2)n}$ depends on $P$, $R$, $Q$
and the delay $h$.\\
Using projection lemma \cite{Ske98}, expression (\ref{proj_ddr2}) is equivalent to (\ref{analyse_dd}).
~\\{\it Remark 1:}
\begin{itemize}
\item There exists another equivalent form of this LMI provided in \cite{Gou06b} and based on quadratic separation.
\item In the same paper, it is shown that for $r=1$, this proposed function (\ref{lyap_ddr}) is equivalent to the main classical results of the literature. Moreover, it is also proved that for $r>1$ conservatism is reduced.
\end{itemize}

\subsection{A first result on synthesis}

Given the analysis condition (\ref{analyse_dd}) and applying the delayed state feedback (\ref{SF}) on system (\ref{forme_can}) (in this subsection, the disturbance is not taken into account), the following proposition is obtained.
 \begin{proposition}
\label{prop_dd2}
 If there exist symmetric positive definite matrices
$P$, $R\in\sf{R}^{n\times n}$, $Q\in\sf{R}^{rn\times rn}$, a matrix
$X\in\sf{R}^{(r+2)n\times n}$, a scalar $h_{m}>0$, an integer $r\geq 1$
and a matrix $K\in\sf{R}^{m\times n}$ such that
\begin{equation}
\label{synthese_dd}  \Gamma+{\bf X}S+S^T{\bf X}^T< \sf{0}
\end{equation}
where $\Gamma$ is defined as (\ref{gamma}) and
\begin{equation}
\label{defS2}
S=\left[\begin{array}{cccc} -\sf{1} & A &\sf{0}_{n\times (r-1)n}&A_d+B{\bf K} \end{array}\right]
\end{equation}
then, system (\ref{forme_can}) can be stabilized for all $h\leq
h_{m}$ for the control law $u(t)=Kx(t)$ (and for $d(t)=0$). \end{proposition}
~\\ \underline{\it Proof}: Considering the system (\ref{forme_can}) with the state feedback (\ref{SF}), the following interconnected system is deduced
\begin{equation}
  \label{interco_sys}
  \dot{x}(t)=Ax(t)+\bar{A}_dx(t-h),
\end{equation}
where $\bar{A}_d=A_d+BK$ and $A$, $A_d$ and $B$ are defined as (\ref{esp_etat}). Then, we can apply the analysis condition (\ref{analyse_dd}) on (\ref{interco_sys}). Using Finsler lemma \cite{Ske98}, there exists a matrix $X\in\sf{R}^{(r+2)n\times n}$ such that if  (\ref{synthese_dd}) is satisfied then (\ref{analyse_dd}) is true. Matrix $X$ is called \textquotedblleft slack variables\textquotedblright which can reduced conservatism and may be interesting for synthesis purpose as well as robust control purpose.
~\\{\it Remark 2:}
\begin{itemize}
\item To solve the synthesis criterion (\ref{synthese_dd}), one has to use a BMI solver.
\item In \cite{Lab07}, a relaxation algorithm is provided with LMI condition to find a stabilizing state feedback.
\end{itemize}

\subsection{Delay dependent state feedback with an integral action}
\begin{figure}
\begin{center}
 \psfrag{p0}[][][0.7]{$p_0$}
 \psfrag{W0}[][][0.7]{$W_0$}
 \psfrag{deltaq}[][][0.7]{$\delta q(t)$}
\includegraphics[height=3cm,width= 5.5cm]{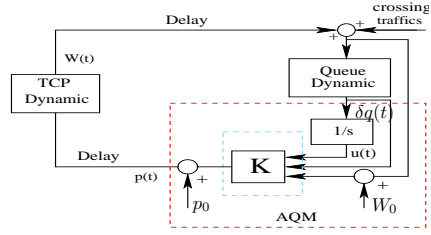}
\caption{Design of an AQM as a dynamic state feedback}
\label{schema_SFI}
\end{center}
\end{figure}

In the previous section, the design of a state feedback control for time delay systems has been exposed. The use of a such controller has been carried out in \cite{Lab07}. However, it appears that in some cases, the queue size is no longer regulated at the desired level (this phenomenon is only observed on the network simulator NS). It thus appears a slight steady state error which can be explain by an inaccuracy of the model. Futhermore, the introduction of non responsive flows like UDP (user datagram protocol) traffics which appear as a disturbance affects the queue size equilibrium and changes the steady state. In order to overcome these problems, the AQM is supplemented with an integral action. The idea is to apply the previously exposed synthesis method over an augmented time delay system composed of the original system (\ref{forme_can}) and an integrator (see figure \ref{schema_SFI}). The augmented system has the following form
\begin{equation}
  \label{aug_sys}
  \dot{z}=
  \left[\begin{array}[h]{cc}
    A&\begin{array}[h]{c} 0\\0\end{array}\\
    \begin{array}[h]{cc}
      0&1
    \end{array}&0
  \end{array}\right]z(t)+
  \left[\begin{array}[h]{cc}
    A_d&\begin{array}[h]{c} 0\\0\end{array}\\
    \begin{array}[h]{cc}
      0&0
    \end{array}&0
  \end{array}\right]z(t-h)
\end{equation}
$+
\left[\begin{array}[h]{c}
  B\\0
\end{array}\right]\delta p(t-h)$\\
with $z^T=[\delta W~~\delta q~~u]^T$ is the extended state variable.
Then, the global control which correspond to our AQM, is a dynamic state feedback
\begin{equation}
  \label{SFI}
  \delta p(t)=K\left[
    \begin{array}{c}
      \delta W(t)\\\delta q(t)\\u(t)
    \end{array}\right]=k_1\delta W(t)+k_2\delta q(t)+k_3\int_0^t \delta q(t)dt.
\end{equation}

In our problem, non modelled crossing traffics $d(t)$ such as UDP based applications are introduced as exogenous signals (see figures \ref{schema_SF} and \ref{schema_SFI}). The queue dynamic is modified as\\
\begin{equation}
  \label{eq_queue_perturbed}
  \dot{q}(t)=\frac{W(t)}{R(t)}N-C+d(t)
\end{equation}
Considering equations (\ref{SFI}), the first equation of (\ref{modelL}) and (\ref{eq_queue_perturbed}), we obtain the transfer function $T(s)$ from the disturbance $D(s)$ to the queue size (about the operating point) $\Delta Q(s)$:
{\small \begin{equation}
   T(s)= \frac{b(s)s}{(s+\frac{1}{R_0})sb(s)+\frac{N}{R_0}\left[\frac{s}{R_0^2C}(1-e^{-hs})+a(s)sk_2+a(s)k_3\right]},
\end{equation}}
 with $a(s)=-\frac{R_0C^2}{2N^2}e^{-hs}$ and $b(s)=s+\frac{N}{R_0^2C}(1+e^{-hs})+a(s)k_1$. It can be easily shown that for a step type disturbance, the queue size still converges to its equilibrium.
\subsection{Estimation of the congestion window}

In these last two parts, a state feedback synthesis has been performed for the congestion control of TCP flows and the management of the router buffer. So far we have considered that the whole state was accessible. However, although the congestion window can be measured in NS (few lines have to be added in the TCP code), it is not the case in reality. That's why, in this paper it is proposed to estimate this latter variable using the aggregate flow incoming to the router buffer.
The sending rate of single TCP source can be approximated by
\begin{equation}
  \label{rate}
  x_i(t)= \frac{W(t)}{R(t)}.
\end{equation}
The above approximation is valid as long as the model does not describe the communication at a finer time scale than few round trip time (see \cite{Low02}). Consequently, the whole incoming rate observed by the router is $x(t)= NW(t)/R(t)$. The measure of the aggregate flow has already been proposed and successfully exploited in \cite{Kun01} and \cite{Kim06} for the realization of the AVQ and a PID type AQM respectively. It is worth noting that queue-based AQMs like RED or PI can be assimilated as output feedbacks according to the queue length. Conversely, AVQ can be viewed as an output feedback with respect to the aggregate flow, belonging thus to the rate-based AQM class.

\section{NS-2 simulations}
\label{simulation}
\begin{figure}[h]
\begin{center}
\vspace*{-1cm}
\hspace*{-0.5cm}
\includegraphics[height=10cm,width= 9cm]{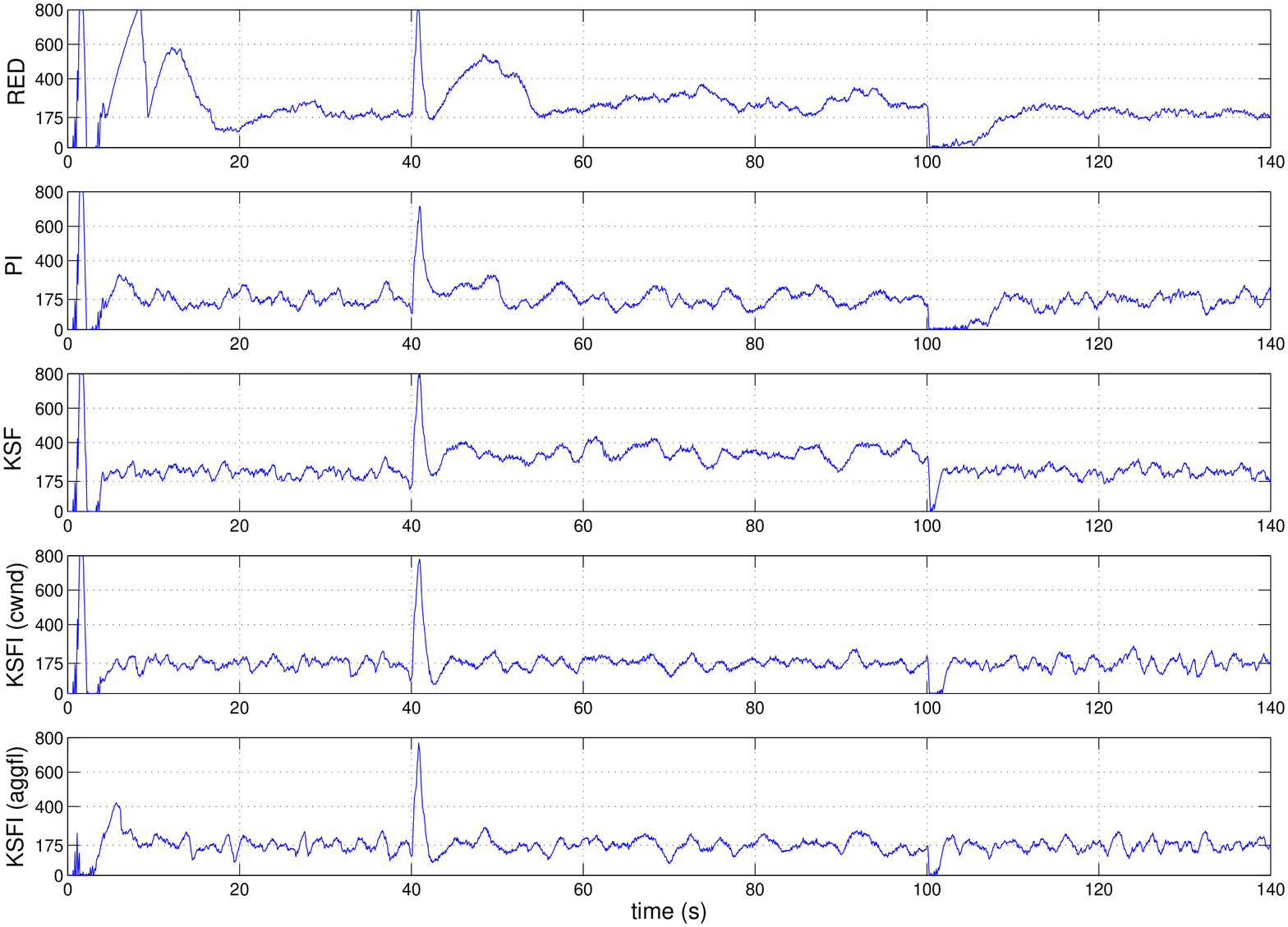}
\vspace*{-1cm}
\caption{Time evolution of the queue length, $AQM=\{RED, PI, K_{SF},
K_{SFI(cwnd)}, K_{SFI(Agg fl)}\}$ under UDP crossing traffic}
\label{courbe_perturbee}
\end{center}
\end{figure}

\begin{figure}[h]
\begin{center}
\hspace*{-0.5cm}
\includegraphics[height=9cm,width= 8cm]{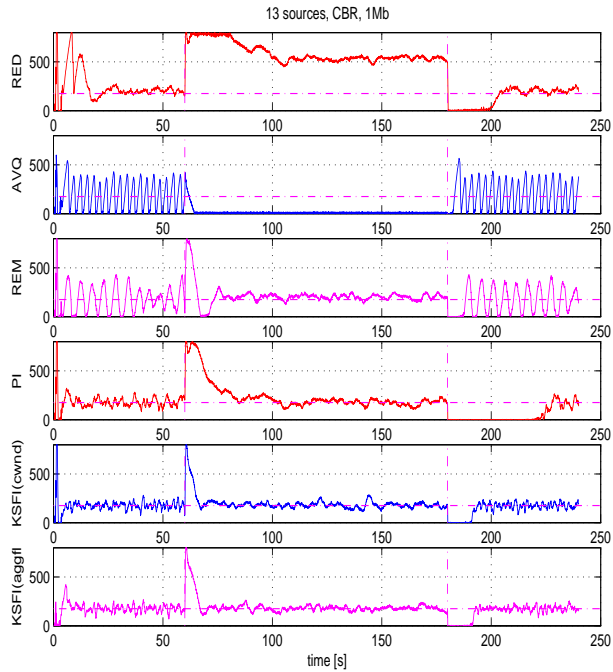}
\caption{Time evolution of the queue length, $AQM=\{RED, AVQ, REM, PI, K_{SFI(cwnd)}, K_{SFI(Agg fl)}\}$ under cross traffic}
\label{courbe_perturbee2}
\end{center}
\end{figure}

As a widely adopted numerical illustration extracted from
\cite{Hol02} (see figure \ref{topology} for the network topology),
consider the case where $q_0=175$ packets, $T_p=0.2$
 second and $C=3750$ packets/s (corresponds to a $15$ Mb/s link with average packet size $500$ bytes). Then, for a load of $N=60$ TCP sessions, we have $W_0=15$ packets, $p_0=0.008$, $R_0=0.246$ seconds. 
According to the synthesis criteria presented in Section \ref{synthesis}, the state feedback matrices
 \begin{equation}
   \label{ex_gains}
   K_{SF}=10^{-3}\left[
     \begin{array}[h]{c}
       -0.2372\\0.0429
     \end{array}\right]\mbox{ and }   K_{SFI}=10^{-4}\left[
     \begin{array}[h]{c}
       0.9385\\0.5717\\0.3559
     \end{array}\right]
 \end{equation}
are calculated for the construction of the control laws (\ref{SF}) and (\ref{SFI}) respectively.\\
We aim at proving the effectiveness of our method using NS-2
\cite{Fal02}, a network simulator widely used in the communication networks
community. Taking values from the previous numerical example, we apply the new
AQM based on a state feedback. The target queue length $q_0$ is
$175$ packets while buffer size is $800$. The average packet length
is $500$ bytes. The default transport protocol is TCP-New Reno
without ECN marking.\\ For the convenience of comparison, we adopt
the same values and network configuration than \cite{Hol02} who
design a PI controller ({\it Proportional-Integral}). This PI is
configured as follow, the coefficients $a$ and $b$ are fixed at
$1.822e-5$ and $1.816e-5$ respectively, the sampling frequency is
$160$Hz. The RED has been also tested using the parametric
configuration recommended in \cite{Hol02}. 
In figure \ref{courbe_perturbee}, simulations are performed under an external perturbation. This latter
is composed of 7 additional sources (CBR applications over UDP protocol) sending 1000
bytes packet length with a 1Mbytes/s throughput between $t=40s$ and
$t=100s$. The two DSF (see figure \ref{courbe_perturbee} for the DSF based on the congestion window and the aggregate flow) regulate faster than others and are able to reject the disturbance swiftly. Conversely, figure \ref{courbe_perturbee} shows the time response of the queue length with a simple state feedback $K_{SF}$ (\ref{SF}) as an AQM. One can note that the queue is stabilized slightly above the desired level (around $200$ pkts). Futhermore, the non reponsive cross traffic affects the steady state.
The table \ref{tableau_stat} summarizes the
benefits of the two $K_{SFI}$ AQMs (according to simulations with UDP cross traffics). Classical statistical
characteristics are calculated during the whole simulation, then
only during the UDP cross traffic and finally after the UDP cross
traffic (come back to steady state). These characteristics are
mean, standard deviation ($Sdt$) and the square of the variation
coefficient ($CV2=(Std/mean)^2$). This latter calculation assess the relative
dispersion of the queue length around its mean.
The mean points out the control precision and the standard deviation shows the ability of the AQM to keep the queue size close to its equilibrium. In table \ref{tableau_stat}, we can observe that $K_{SFI} (cwnd)$ maintains a very good control on the buffer queue during the whole simulation. Even though $K_{SFI} (aggfl)$ is slightly slower than the previous one, statistics (Std and CV2) show again a good regulation. Although PI reject the perturbation quite fast, extensive fluctuations appear during the steady state. To conclude, the two DSF are efficient AQMs which provide the best precision and are able to regulate faster and closer to the mean compared to others AQMs.
\begin{table}[h]
\begin{center}
{\small
\begin{tabular}{|c|c|c|c|c|c|c|} \hline
\!AQMs\! & RED & PI & $K_{SF}$ & \begin{tabular}{c}\!\!\!\!$K_{SFI}$\!\!\!\!\\\!\!\!\!(cwnd)\!\!\!\!\end{tabular} & \begin{tabular}{c}\!\!\!\!$K_{SFI}$\!\!\!\!\\\!\!\!\!(aggfl)\!\!\!\!\end{tabular}& \\ \hline\hline
\!Mean\! & 235.7 & 176.7 & 263.9 &\!\!\!\! 175.9\!\!\!\! &\!\!\!\! 175.5\!\!\!\!& B\\ \hline
\!Sdt\! & 112.40 & 71.19 & 78.59 &\!\!\!\! 54.57\!\!\!\! &\!\!\!\! 63.64\!\!\!\!& B\\ \hline
\!CV2\! & 0.227 & 0.162 & 0.088 &\!\!\!\! 0.096\!\!\!\! &\!\!\!\! 0.131\!\!\!\!& B\\ \hline\hline
\!Mean\! & 270.3 & 178.3 & 338.0 &\!\!\!\! 173.4\!\!\!\! &\!\!\!\! 175.6\!\!\!\!& D\\ \hline
\!Sdt\! & 57.39 & 40.42 & 41.21 &\!\!\!\! 35.08\!\!\!\! &\!\!\!\! 28.32\!\!\!\!& D\\ \hline
\!CV2\! & 0.045 & 0.051 & 0.014 &\!\!\!\! 0.040\!\!\!\! &\!\!\!\! 0.026\!\!\!\!& D\\ \hline\hline
\!Mean\! & 201.4 & 177.8 & 236.0 &\!\!\!\! 176.2\!\!\!\! & \!\!\!\!174.8\!\!\!& A\\ \hline
\!Sdt\! & 22.24 & 36.64 & 30.48 &\!\!\!\! 30.22\!\!\!\! &\!\!\!\! 32.64\!\!\!\!& A\\ \hline
\!CV2\! & 0.012 & 0.042 & 0.016 &\!\!\!\! 0.029\!\!\!\! &\!\!\!\! 0.034\!\!\!\!& A\\ \hline
\end{tabular}}
\end{center}
\vspace*{0.4cm}
\caption{Statistical characteristics for different AQMs (units are pkts) at different periods (B, D and A: before, during and after CBR applications)}
\label{tableau_stat}
\end{table}
To complete our simulation, we propose another NS-2 simulation between different AQMs: REM, AVQ, RED, PI, KSFI(cwnd) and KSFI(aggfl). We consider different levels of CBR cross traffics (13 sources, 1Mb). RED, REM, PI and AVQ are fixed to same values as in \cite{toto}, which is a performance analysis of AQM under DoS attacks. The additional sources are sending 1000 bytes packet length with a 1Mbytes/s throughput between $t=60s$ and $t=180s$. The simulation is illustrated in the figure \ref{courbe_perturbee2}. In these last two cases, one can imagine that AQMs could detect cross traffics or traffic anomalies. Moreover, $AQM=\{K_{SFI(cwnd)}, K_{SFI(Agg fl)}\}$ still have well behaviours under cross traffic conditions.
\section{Conclusion}
\label{conclusion}
In this preliminary work, we have proposed the design of an
 AQM for the congestion control in communications networks.
The developed AQM has been constructed using a dynamic state
feedback control law. An integral action has been added to reject
the steady state error in spite of disturbance, $d(t)$ (cross traffic). Finally, the AQM has
been validated using NS simulator. Future work consist in the improvement
about control laws (theoritical part) extended to a greater network using a decentralized approach to reduce the weakness of this method on one side and validation on emulation platform (experimental part) on the other side.
\bibliographystyle{plain}
\bibliography{mabiblio}

\end{document}